\documentclass[a4paper]{PoS}
\usepackage[utf8]{inputenc}
\usepackage{tikz}
\usepackage{bbold}
\usepackage{amsmath}
\usepackage{mcite}
\usepackage{cite}

\title{
	\begin{flushright}\begin{small}
			{\it NT@UW-18-14~\\
	CP3-Origins-2018-040 DNRF90\\} \vspace{0.5cm}
		\end{small}
	\end{flushright}
	Stabilising complex Langevin simulations}

\ShortTitle{Stabilising complex Langevin simulations}

\author{\speaker{Felipe Attanasio}\\
Department of Physics, University of Washington, Box 351560, Seattle, WA 98195, USA\\
E-mail: \email{pyfelipe@uw.edu}}

\author{Benjamin J\"ager\\
CP3-Origins \& Danish Institute for Advanced Study, Department of Mathematics and Computer Science, University of Southern Denmark, 5230 Odense M, Denmark\\
E-mail: \email{jaeger@cp3.sdu.dk}}

\abstract{We present updated results of dynamic stabilisation (DS) applied to complex Langevin simulations of QCD in the heavy-dense limit and with staggered quarks.
We show that DS is able to keep the unitarity norm sufficiently small, which leads to excellent agreement with Monte-Carlo simulations, when the latter is applicable.}

\FullConference{The 36th Annual International Symposium on Lattice Field Theory - LATTICE2018\\
		22-28 July, 2018\\
		Michigan State University, East Lansing, Michigan, USA.}

\begin{document}

\section{Introduction}
The phase diagram of strongly interacting matter at finite temperature $T$ and baryon chemical potential $\mu$ is a very active field of research.
Knowledge of the different phases has important applications for determining the QCD equation of state, as well as understanding the phase transitions that quarks and gluons go through.
There are planned (FAIR, NICA) and ongoing experimental investigations (RHIC, LHC) of QCD under different thermodymic conditions.

The theoretical study of the QCD phase diagram requires the use of non-perturbative methods, since perturbation theory is only applicable at high $T$ or $\mu$, where quarks are asymptotically free.
Traditionally, QCD is simulated via Monte Carlo methods in Euclidean spacetime.
The addition of a baryon, or quark, chemical potential makes the action complex, leading to a complex probablity weight -- this is known as the \textit{sign problem}.
In situations where the sign problem is mild, i.e., the phase of the weight does not change much, Taylor expansions or methods such as reweighting can be applied. For a review, see~\cite{deForcrand:2010ys}.
However, when $\mu / T \gtrsim 1$ an exponentially hard overlap problem prevents the obtention of reliable results.

\section{Complex Langevin}
One promising technique to circumvent the sign problem is the complex Langevin method.
This method is based on stochastic quantisation~\cite{Parisi:1980ys}, where the dynamical variables evolve in a fictitious time dimension $\theta$ according to a Langevin equation.
Quantum expectation values are recovered as averages over $\theta$ after the system reaches its stationary state.

Specialising for SU($3$) gauge fields, the Langevin equation reads
\begin{equation}
	U_{x\mu}(\theta+\varepsilon) = \exp\left[X_{x\mu}\right] U_{x\mu}(\theta)\,,
\end{equation}
\begin{equation}
	X_{x\mu} = i\lambda^a(-\varepsilon D^a_{x\mu}S\left[U(\theta)\right] + \sqrt{\varepsilon} \, \eta^a_{x\mu}(\theta))\,,
\end{equation}
with $U_{x\mu}(\theta)$ being the gauge links at Langevin time $\theta$, $X_{x\mu}$ the Langevin drift, $\lambda^a$ are the Gell-Mann matrices, $\varepsilon$ is the step size, which is chosen adaptively~\cite{Aarts:2009dg}, $\eta^a_{x\mu}$ are white noise fields satisfying
\begin{equation}
	\langle\eta^a_{x\mu}\rangle = 0\,, \quad \langle \eta^a_{x\mu} \eta^b_{y\nu} \rangle = 2 \delta^{ab} \delta_{xy} \delta_{\mu\nu}\,,
\end{equation}
$S$ is the QCD action and $D^a_{x\mu}$ is defined as
\begin{equation}
	D^a_{x\mu} f(U) = \left.\frac{\partial}{\partial \alpha} f(e^{i\alpha\lambda^a} U_{x\mu})\right|_{\alpha=0}\,.
\end{equation}

The complex character of the method comes into play, since $S$ is complex, by allowing the gauge links to take values in the complex extension of SU($3$), namely the group SL($3,\mathbb{C}$)~\cite{Aarts:2008rr,Aarts:2008wh,Aarts:2010gr,Aarts:2011zn}.
To ensure convergence, the action and observables must be holomorphic functions~\cite{Aarts:2009uq}, and therefore the substitution $U^\dagger \to U^{-1}$ is necessary, since they are the same on SU($3$).

\begin{figure}
	\centering
\begin{minipage}{0.48\textwidth}
	\centering
	\includegraphics[width=1.0\textwidth]{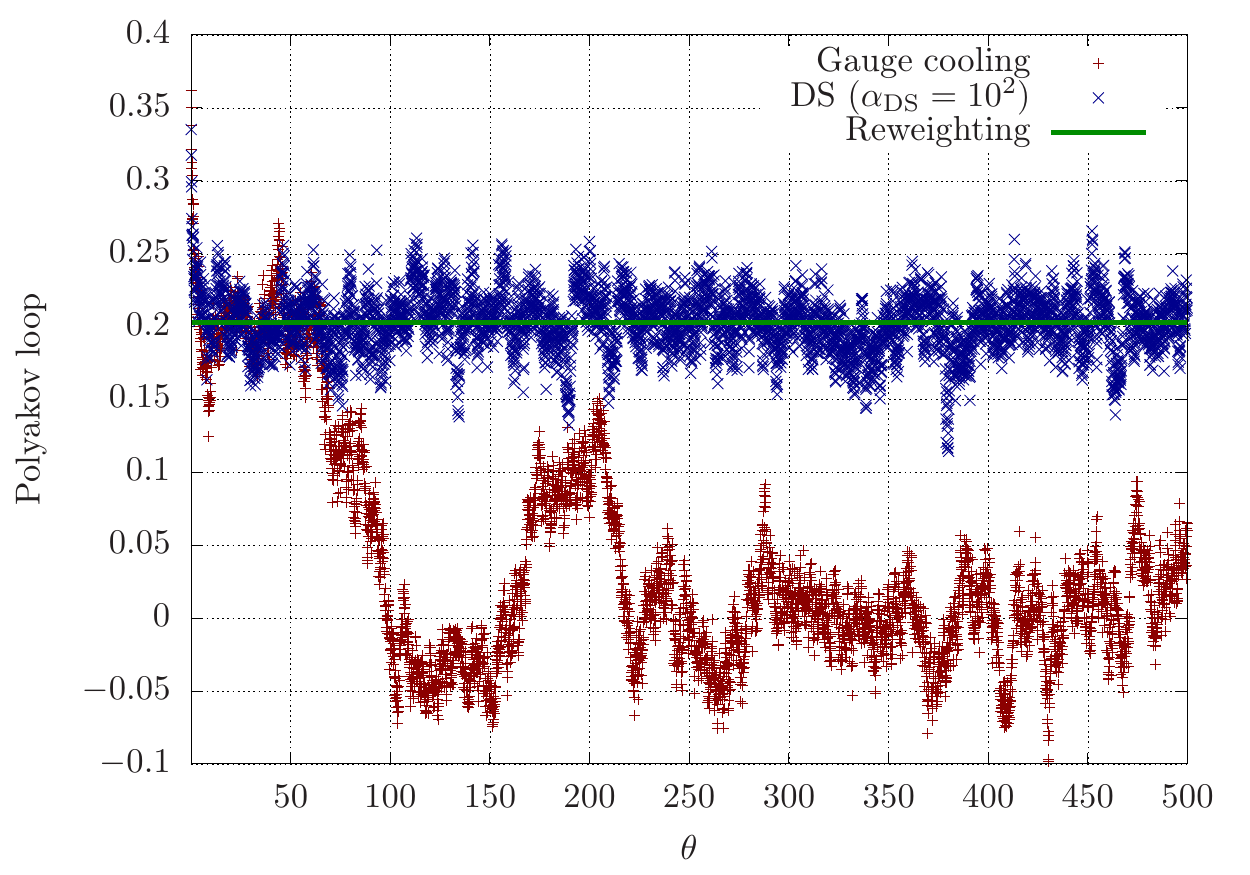}
\end{minipage}
\begin{minipage}{0.48\textwidth}
	\centering
	\includegraphics[width=1.0\textwidth]{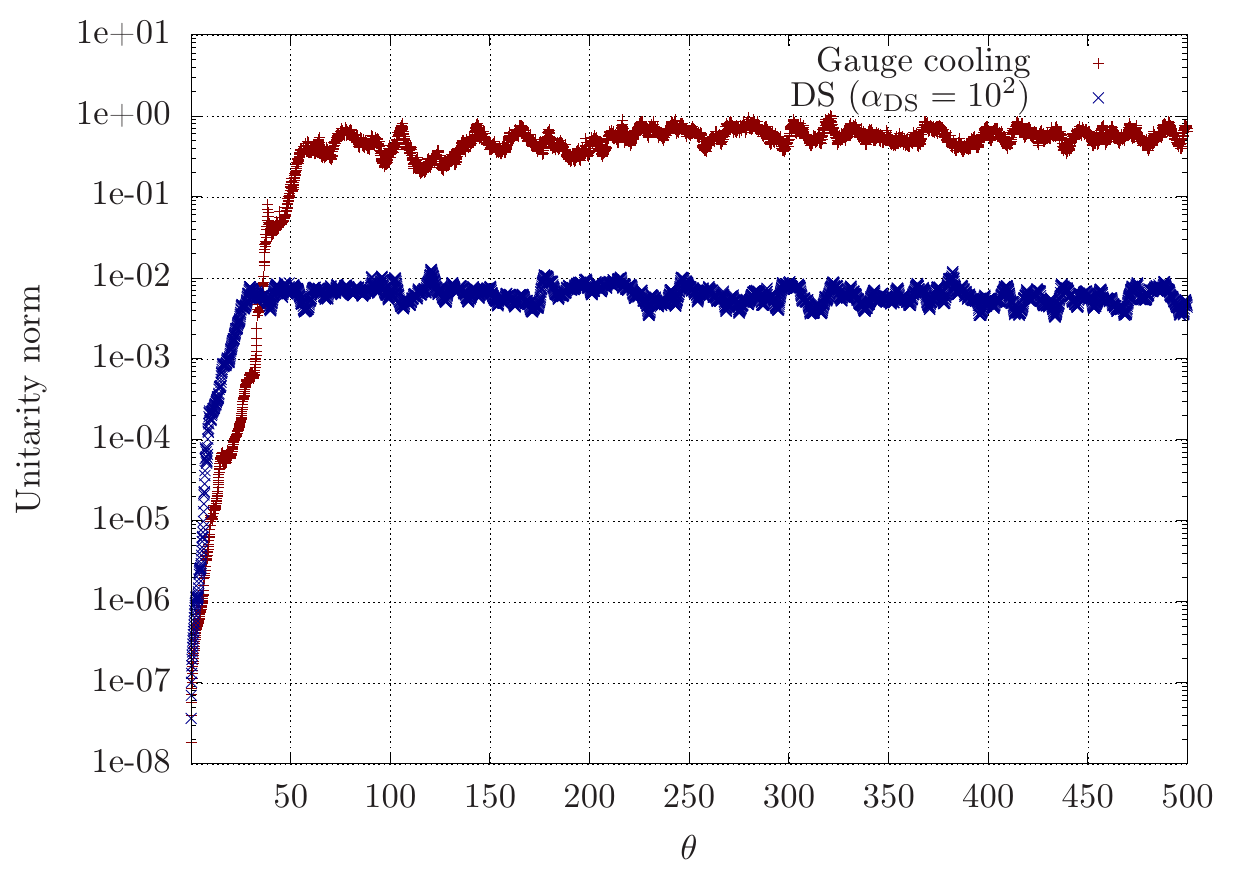}
\end{minipage}
\caption{\label{fig.DS.GC.RW}Left: Polyakov loop as a function of Langevin time for different simulations. Right: Langevin time history of the unitarity norm. Instabilities for the simulation with only gauge cooling start when the unitarity norm reaches $\mathcal{O}(0.1)$.}
\end{figure}

An issue arises from the fact that SL($3,\mathbb{C}$) is not compact: the simulation might follow an unstable trajectory and converge to a wrong limit~\cite{Ambjorn:1985iw,Ambjorn:1986fz,Aarts:2010aq}.
We monitor the distance from SU($3$), given by the unitarity norm
\begin{equation}
	d = \frac{1}{3\Omega} \sum_{x,\mu} \text{Tr} \left[ U_{x,\mu} U^\dagger_{x,\mu} - \mathbb{1} \right] \geq 0\,,
\end{equation}
with $\Omega = N_s^3 \times N_\tau$ being the lattice 4-volume, and use gauge transformations to reduce it, with the gauge cooling technique~\cite{Seiler:2012wz,Aarts:2013uxa,Aarts:2015hnb}.
Gauge cooling, despite being necessary, is not sufficient to keep $d$ fully under control~\cite{Aarts:2016qrv}.
In order to remedy this situation, we have proposed the method of Dynamic Stabilisation~\cite{Attanasio:2018rtq}.
It consists of a modification of the Langevin drift,
\begin{equation}
	X_{x\mu} \to X_{x\mu} + i \alpha_{\mathrm{DS}} \lambda^a M^a_x\,,
\end{equation}
such that i) $d$ does not exceed a given threshold; ii) the SU($3$) part of the Langevin drift is not changed.
One possible implementation is
\begin{equation}
	M^a_x = i b^a_x \left( \sum_c b^c_x b^c_x \right)^3\,, \quad
	b^a_x = \text{Tr} \left[ \lambda^a \sum_\nu U_{x\nu} U^\dagger_{x\nu} \right]\,.
\end{equation}
A comparison between simulations using gauge cooling, gauge cooling and dynamic stabilisation, and reweighting in QCD in the limit of heavy quarks (HDQCD)~\cite{Bender:1992gn,Aarts:2008rr} is seen in fig.~\ref{fig.DS.GC.RW}(left).
On the right hand side, we show the unitarity norm $d$.
Note that the region where the Polyakov loop converges to the wrong limit coincindes with $d$ being of order $0.1$.

\section{Testing dynamic stabilisation in HDQCD}
In the heavy-dense approximation of QCD (HDQCD), quarks can only evolve in the Euclidean time direction.
This greatly simplifies the fermion determinant, but preserves a silver blaze problem at $T=0$ and the sign problem at real chemical potential.
The QCD action with fermions integrated out reads
\begin{equation}
	S = S_{\mathrm{YM}} - \ln \det M(U,\mu)\,,
\end{equation}
with $S_{\mathrm{YM}}$ being the Wilson gauge action and the fermion determinant,
\begin{align}
	\det M(U,\mu) = &\prod_{\vec{x}} \left\{ \det \left[ 1 + \left(2\kappa e^\mu \right)^{N_\tau} \mathcal{P}_{\vec{x}} \right]^2 \det \left[ 1 + \left(2\kappa e^{-\mu} \right)^{N_\tau} \mathcal{P}_{\vec{x}}^{-1} \right]^2 \right\}\,,
\end{align}
being a function of the Polyakov loop and its inverse,
\begin{equation}
	\mathcal{P}_{\vec{x}} = \prod_\tau U_4(\vec{x},\tau)\,,\quad \mathcal{P}^{-1}_{\vec{x}} = \prod^0_{\tau=N_\tau-1} U^{-1}_{(\vec{x},\hat{4})}.
\end{equation}

\begin{figure}
	\centering
	\includegraphics[scale=0.65]{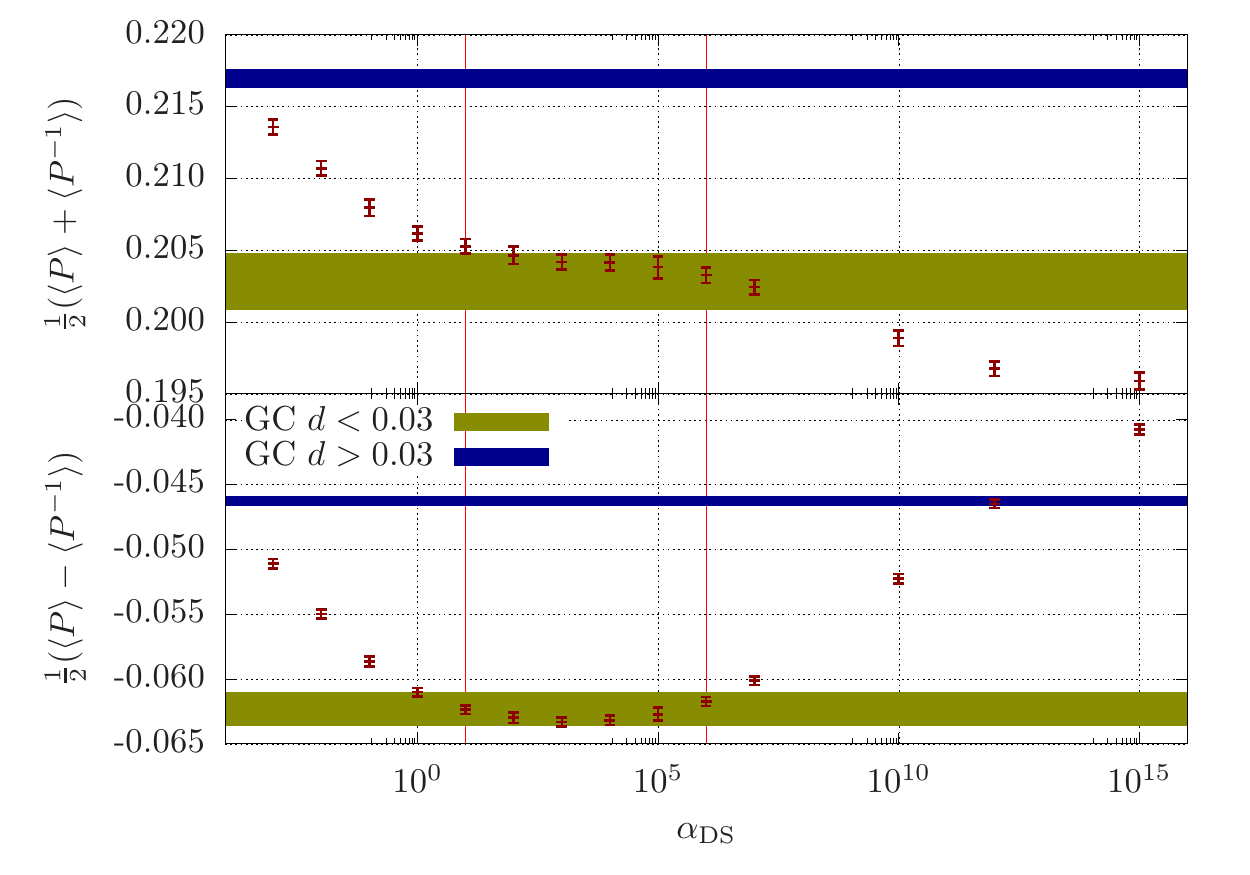}
	\caption{\label{fig.poly.alpha.scan}Real and imaginary parts of the Polyakov loop as functions of $\alpha_{\mathrm{DS}}$. The bands are results from gauge cooling (GC) with and without a cut in the unitarity norm of $d<0.03$.}
\end{figure}

In fig.~\ref{fig.poly.alpha.scan} we analyse the dependency of the Polyakov loop on $\alpha_{\mathrm{DS}}$ for a situation where HDQCD exhibits a severe sign problem, with volume $\Omega=8^3 \times 20$, $\beta=5.8$ $\kappa=0.04$ and $\mu=2.45$.
We compare the results with simulations using only gauge cooling.
In one case, we included all points in the analysis (marked as $d>0.03$), and in the other we stopped the analysis when the unitarity norm became larger then $0.03$.
Both situations are shown as coloured bands in fig.~\ref{fig.poly.alpha.scan}.
We found a wide region in the DS parameter where both real and imaginary parts of the Polyakov loop agree with the gauge cooling results restricted to $d<0.03$.

\begin{figure}
	\centering
\begin{minipage}{0.48\textwidth}
	\centering
	\includegraphics[width=1.0\textwidth]{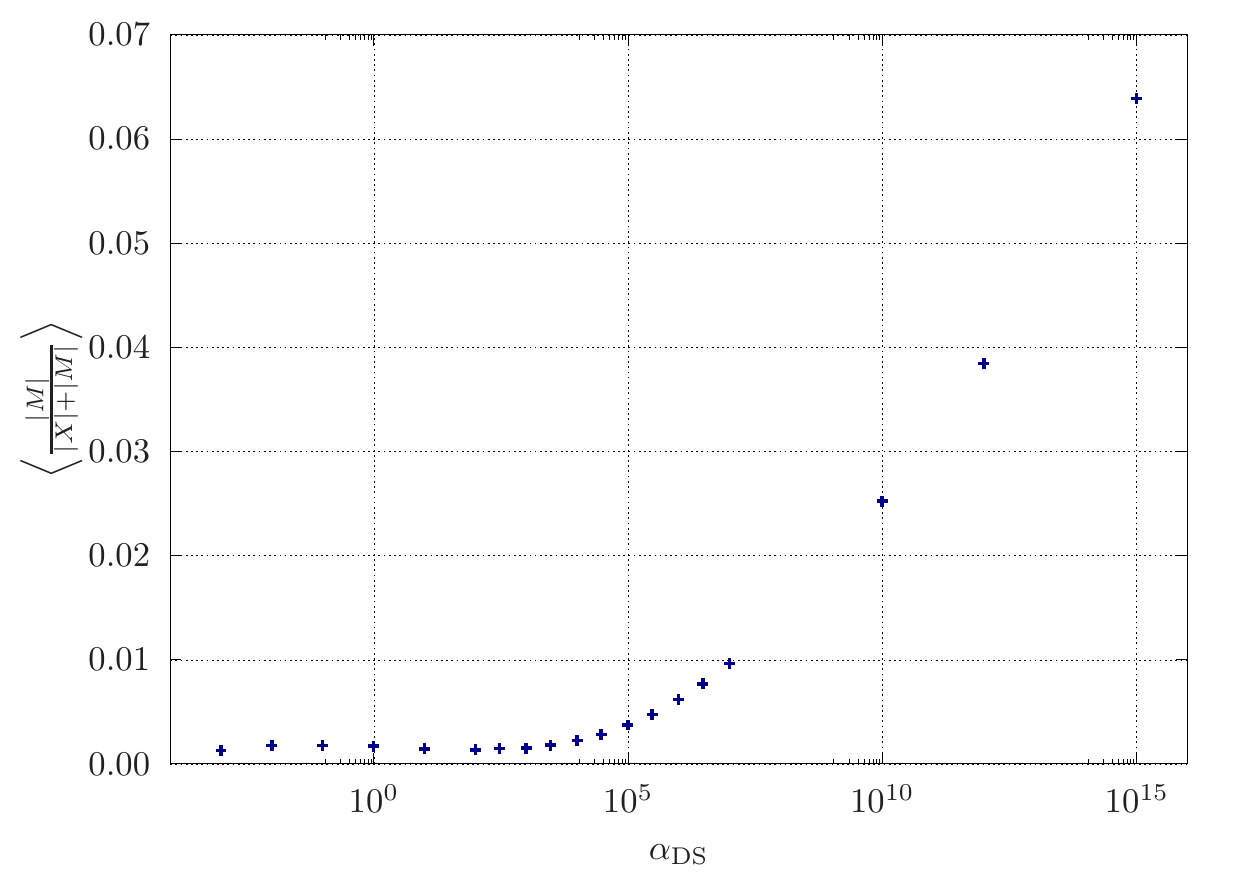}
\end{minipage}
\begin{minipage}{0.48\textwidth}
	\centering
	\includegraphics[width=1.0\textwidth]{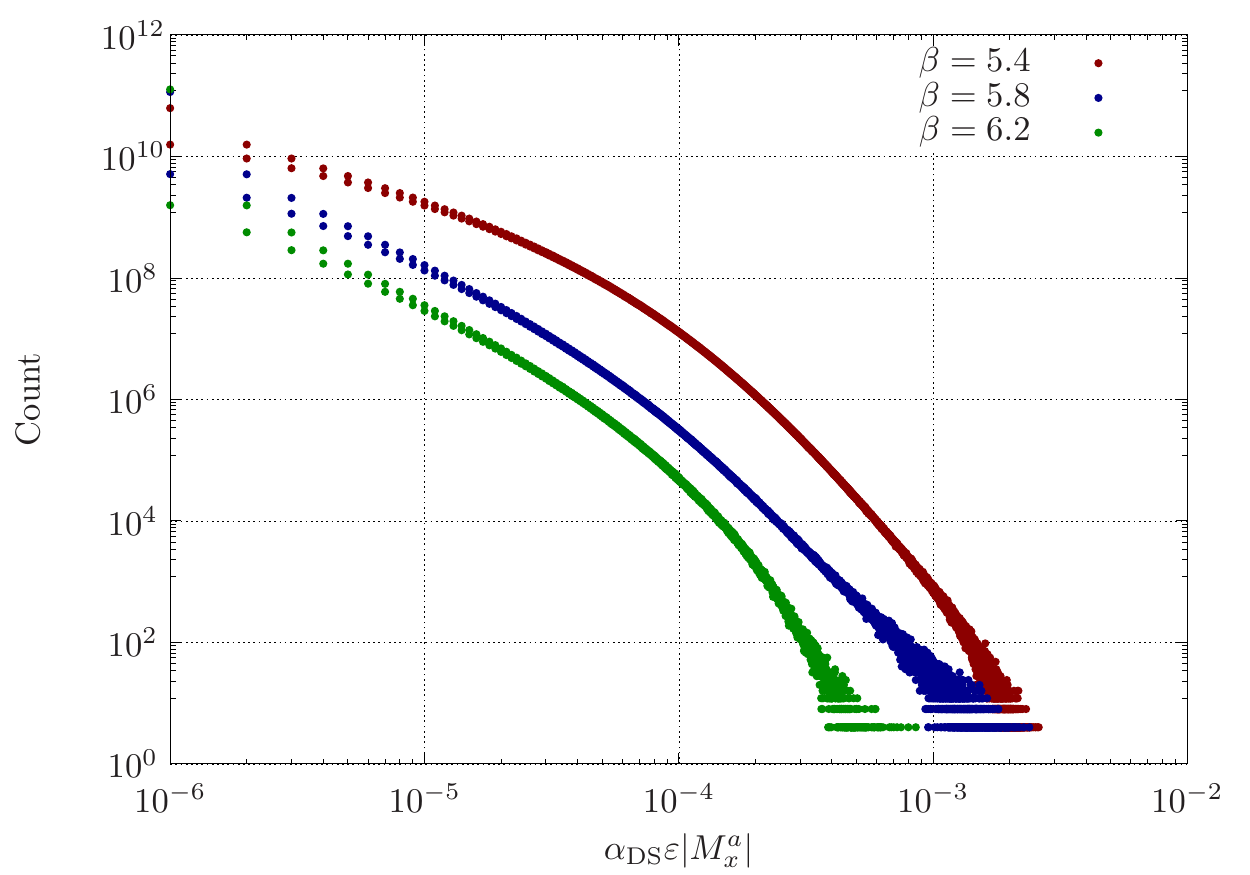}
\end{minipage}
\caption{\label{fig.HD.ratios}Left: Ratio of the DS and total drifts as a function of $\alpha_{\mathrm{DS}}$. Right: Histograms of the DS drift for different values of the inverse coupling.}
\end{figure}

In fig.~\ref{fig.HD.ratios} we investigate the contribution of the DS term in the same situation studied above.
On the left panel, we consider the ratio of the DS to the total drift
As the drifts are complex, we study their absolute values.
For the total drift, we consider $|X|$ and $|M|$ separately, in order to have a ratio between $0$ and $1$.
The results for the average of $|M|/(|X|+|M|)$ show that the DS contribution is never above $7\%$ of the total drift even for the largest value of $\alpha_{\mathrm{DS}}$ considered.
The panel on the right of fig.~\ref{fig.HD.ratios} shows histograms of the DS drift for different values of $\beta$.
We can see that the DS contribution becomes smaller at finer lattices.

\section{Staggered quarks}
We have employed the complex Langevin method, augmented with gauge cooling and dynamic stabilisation to study QCD with dynamical fermions.
The Langevin drift has the form
\begin{equation}
	X_{x\mu} = \lambda^a \left( D^a_{x\mu} S_{\mathrm{YM}} - \text{Tr}\left[ M^{-1} D^a_{x\mu} M \right] + i \alpha_{\mathrm{DS}} M^a_x \right)\,,
\end{equation}
with $M$ being the fermion matrix.
In order to evaluate the term stemming from the quark action, we employ a bilinear noise scheme for the trace~\cite{Batrouni:1985jn}, and conjugate gradient for $M^{-1}$.
Potential issues caused by poles in $M^{-1}$ have been investigated in~\cite{Aarts:2017vrv}.
One consequence of using the bilinear noise scheme is that even for $\mu=0$ the drift is real only on average, and therefore an imaginary part can appear during the simulation.
This can cause the simulation to diverge.
An alternative, exact way of evaluating the fermion contribution has been studied in~\cite{Bloch:2017jzi}.
We study whether DS is able to allow for convergence to the right limit to happen, without projecting the system back to SU($3$).
This convergence is checked against results from Hybrid Monte-Carlo (HMC) simulations\footnote{We thank Philippe de Forcrand for providing us with these results.}.
The simulations were carried out with four flavours of staggered quarks, volumes of $6^4$, $8^4$, $10^4$ and $12^4$, inverse coupling $\beta=5.6$ and mass $m=0.025$.

\begin{figure}
	\centering
\begin{minipage}{0.48\textwidth}
	\centering
	\includegraphics[scale=0.52]{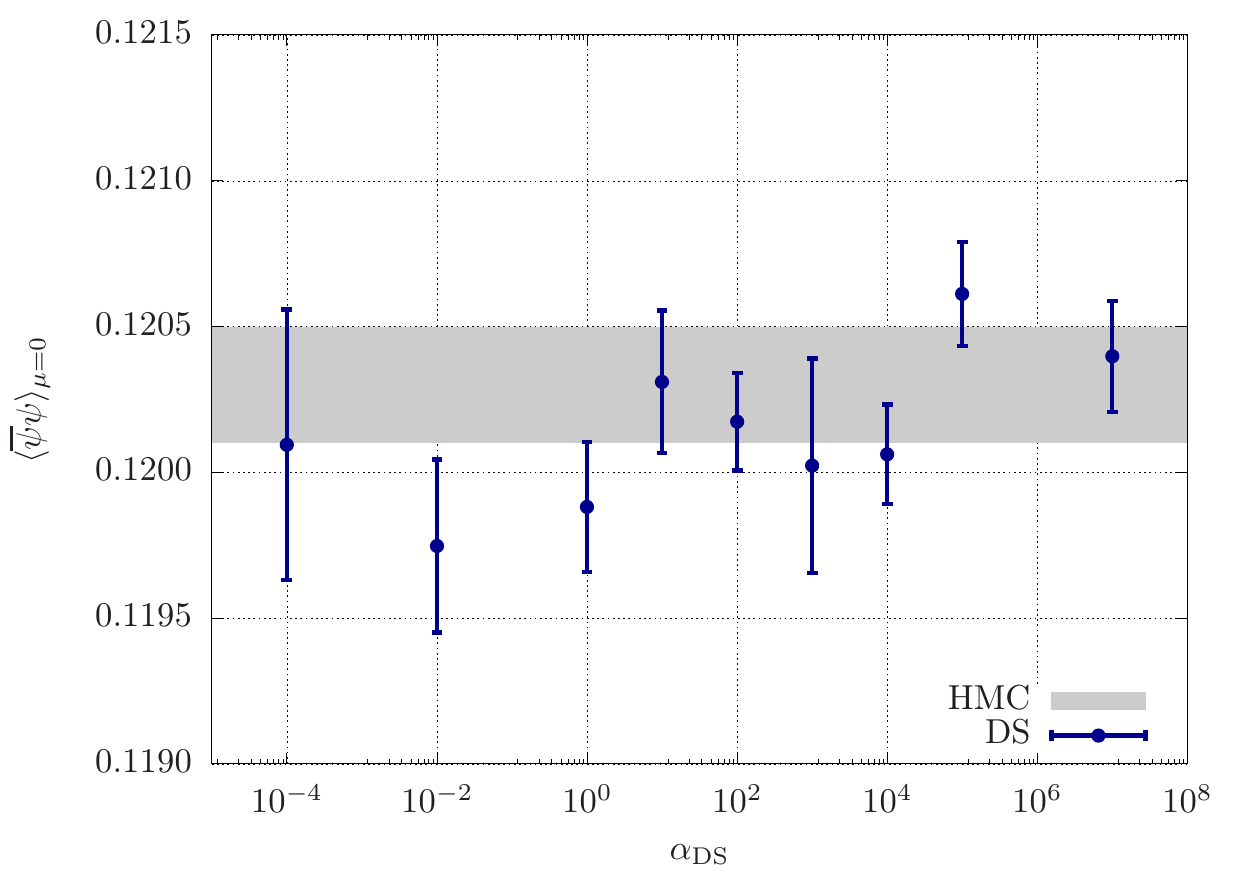}
\end{minipage}
\begin{minipage}{0.48\textwidth}
	\centering
	\includegraphics[scale=0.52]{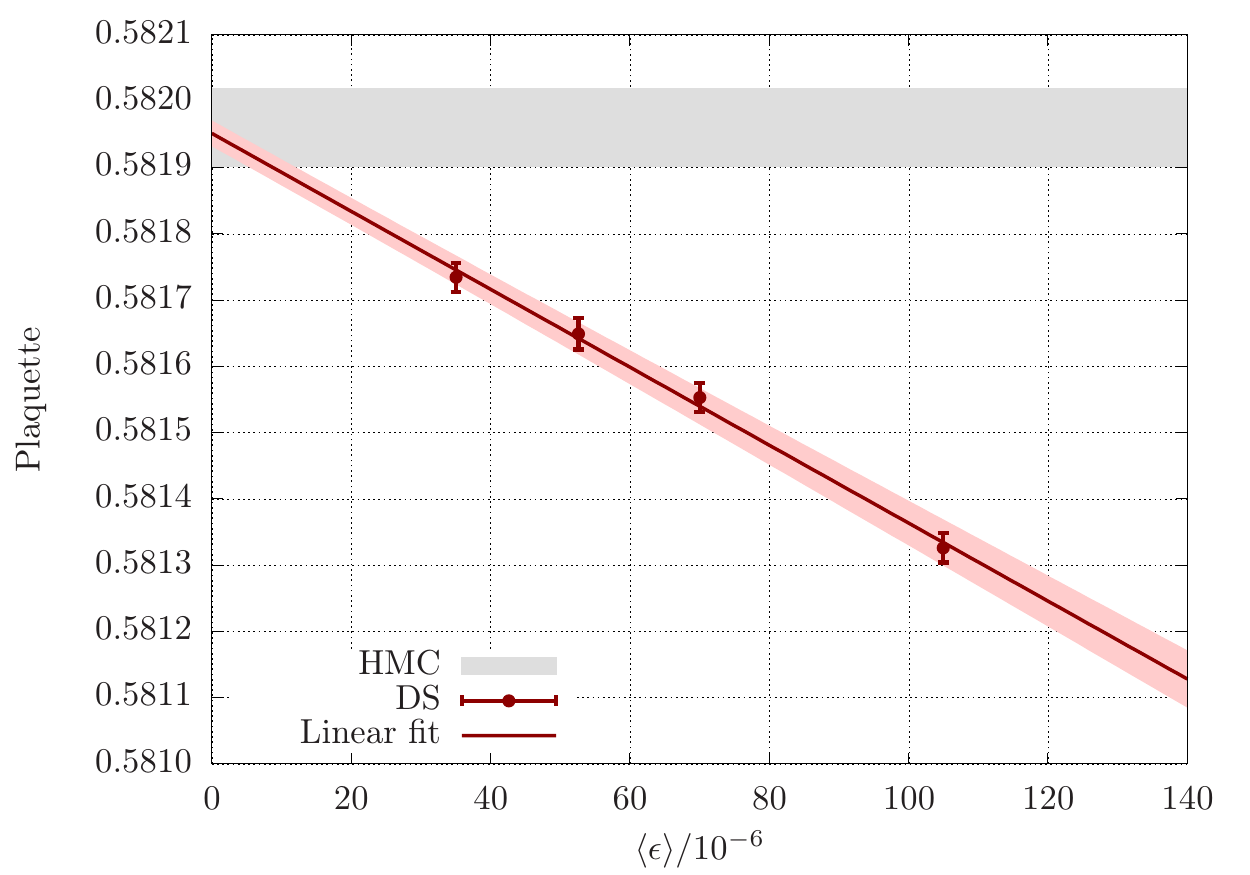}
\end{minipage}
	\caption{\label{fig.cc.vol6}Left: Average chiral condensate as function of $\alpha_{\mathrm{DS}}$ condensate at $\mu=0$ and volume of $6^4$.
	Right: Extrapolation of the Langevin step size to zero for the average plaquette at $\mu=0$ and $V=12^4$.}
\end{figure}

Figure~\ref{fig.cc.vol6} (left) shows the chiral condensate as a function of $\alpha_{\mathrm{DS}}$ resulting from Langevin simulations, for a volume of $6^4$.
The grey band indicates the result from HMC.
Good agreement is observed for all $\alpha_{\mathrm{DS}}$, despite the finite Langevin step size.
The extrapolation to zero Langevin step size can be seen in fig.~\ref{fig.cc.vol6} (right), for the volume of $12^4$.
Since we used a first order discretisation scheme for the Langevin evolution, the step size corrections are linear in $\epsilon$~\cite{Damgaard:1987rr}.
This is confirmed by the linear fit provided.
Excellent agreement is seen between the error bands from Langevin and HMC simulations.

Our results for the average plaquette and chiral condensate, extrapolated for zero step size, for all studied volumes is shown in table~\ref{tb.stag}, as well as the results from the HMC simulations.
Excellent agreement has been found for both observables in all four volumes considered.
\begin{table}
\caption{\label{tb.stag}Average plaquette and chiral condensate obtained from Langevin and HMC simulations of four flavours of staggered fermions at $\beta=5.6$, $m=0.025$ and $\mu=0$ in four different lattice volumes.}
\centerline{
\begin{tabular}{ccccc}
	\hline\noalign{\smallskip}
	 & \multicolumn{2}{c}{$\overline{\psi} \psi$} & \multicolumn{2}{c}{Plaquette} \\
	 Volume & HMC & Langevin & HMC & Langevin \\
	\noalign{\smallskip}\hline\noalign{\smallskip}
$6^4$  & $0.1203(3)$ & $0.1204(2)$ & $0.58246(8)$ & $0.582452(4)$ \\
$8^4$  & $0.1316(3)$ & $0.1319(2)$ & $0.58219(4)$ & $0.582196(1)$ \\
$10^4$ & $0.1372(3)$ & $0.1370(6)$ & $0.58200(5)$ & $0.58201(4)$  \\
$12^4$ & $0.1414(4)$ & $0.1409(3)$ & $0.58196(6)$ & $0.58195(2)$  \\
\noalign{\smallskip}\hline
\end{tabular}
}
\end{table}

\begin{figure}
	\centering
	\includegraphics[scale=0.65]{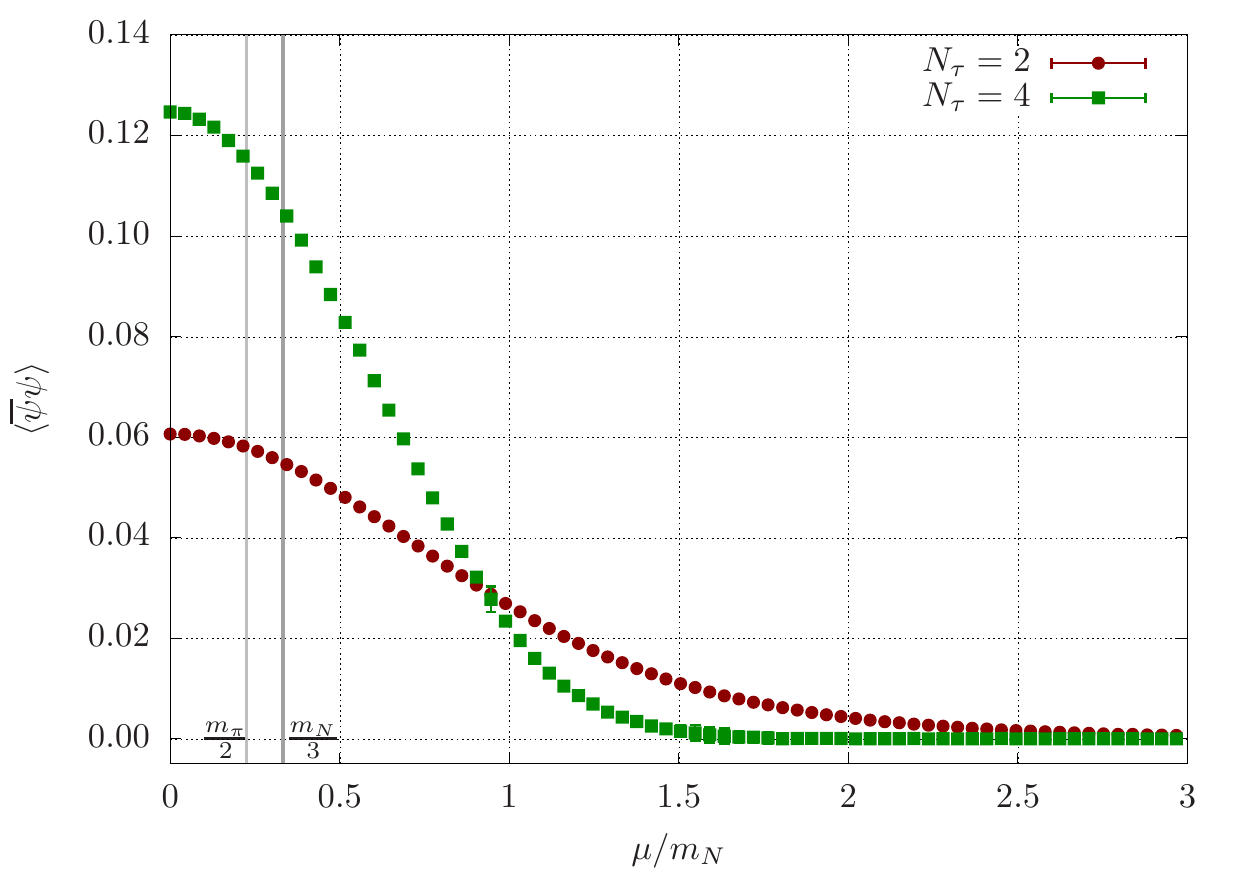}
	\caption{\label{fig.cc.finite.temp}Chiral condensate as function of the chemical potential, in units of the nucleon mass. Lines of pion (left) and baryon (right) condensation are indicated.}
\end{figure}

We show in fig. \ref{fig.cc.finite.temp} a qualitative study of staggered fermions at finite $\mu$, with $N_f=2$, $\beta=5.6$, $V=12^3$ and mass $m=0.025$ for different temperatures, $N_\tau=2$ and $4$.
The pion and nucleon masses are $m_\pi \approx 0.42$ and $m_N \approx 0.93$, respectively \cite{Bitar:1993rk}.
At high temperatures, the inversion of the fermion matrix is numerically cheap and converges quickly.
\vspace{-0.5cm}

\section{Summary}
We report on results of our method of dynamic stabilisation applied to complex Langevin simulations.
In situations where the distance from the unitary manifold exceeds $\mathcal{O}(0.1)$ instabilities have been observed to cause simulations to converge to wrong limits.
Dynamic stabilisation (DS) has been constructed to act as force added to the Langevin drift that keeps the system from exploring too far in the SL($3, \mathbb{C}$) manifold.

We have shown that by employing DS and tuning its control parameter, it is possible to get correct convergence for QCD simulations both in the limit of heavy quarks and light quarks at zero chemical potential.
In the case of HDQCD, we have shown that DS only adds a small contribution to the Langevin drift.
For QCD with fully dynamical quarks, we have found excellent agreement with HMC results at zero chemical potential after extrapolating the Langevin step size to zero.

\acknowledgments
We would like to thank Gert Aarts, D\'enes Sexty, Erhard Seiler and Ion-Olimpiu Stamatescu for
invaluable discussions and collaboration. We are indebted to Philippe de Forcrand for providing
us with the HMC results for staggered fermions. We are grateful for the computing resources made
available by HPC Wales. This work was facilitated though the use of advanced computational,
storage, and networking infrastructure provided by the Hyak supercomputer system at the University of Washington.
The work of FA was supported by US DOE Grant No. DE-FG02-97ER-41014.

\bibliographystyle{style}
\bibliography{ref}

\end{document}